 \documentclass[final,5p,times,twocolumn,authoryear]{elsarticle}

\usepackage{amssymb}
\usepackage{lipsum}
\usepackage{url}

\usepackage{xpatch}
\usepackage{listings} 

\usepackage[pdf]{graphviz}
\usepackage[dvipsnames,svgnames]{xcolor}
\usepackage{listings}

\definecolor{codegray}{rgb}{0.5,0.5,0.5}
\definecolor{backcolour}{rgb}{0.95,0.95,0.92}
\lstdefinestyle{defaultcodestyle}{
    backgroundcolor=\color{backcolour},
    commentstyle=\color{codegreen},
    keywordstyle=\color{magenta},
    numberstyle=\tiny\color{codegray},
    stringstyle=\color{codepurple},
    basicstyle=\ttfamily\tiny,
    breakatwhitespace=false,
    breaklines=false,
    captionpos=b,
    keepspaces=true,
    numbers=left,
    numbersep=5pt,
    showspaces=false,
    showstringspaces=false,
    showtabs=false,
    tabsize=2,
    xleftmargin=0.5cm,
}
\makeatletter
\newcommand*{\addFileDependency}[1]{
  \typeout{(#1)}
  \@addtofilelist{#1}
  \IfFileExists{#1}{}{\typeout{No file #1.}}
}
\makeatother
\xpretocmd{\digraph}{\addFileDependency{#2.dot}}{}{}

\newcommand{\karabo}{Karabo}
\newcommand{\oskar}{\texttt{OSKAR}}
\newcommand{\rascil}{\texttt{RASCIL}}
\newcommand{\wsclean}{\texttt{WSCLEAN}}
\newcommand{\skymodel}{\texttt{SkyModel}}
\newcommand{\telescope}{\texttt{Telescope}}

\newcommand{\image}{\texttt{Image}}
\newcommand{\imager}{\texttt{Imager}}
\newcommand{\software}{\texttt{Software}}
\newcommand{\healpix}{\texttt{healpix}}
\newcommand{\imagemosaicker}{\texttt{ImageMosaicker}}

\journal{Astronomy $\&$ Computing}

\begin{document}

\begin{frontmatter}


\title{Karabo: A versatile SKA Observation Simulation Framework}



\author[first,second]{Rohit Sharma}
\affiliation[first]{organization={Space, Planetary \& Astronomical Sciences \& Engineering (SPASE), Indian Institute of Technology Kanpur},
            addressline={Kalyanpur}, 
            city={Kanpur},
            postcode={208016}, 
            state={Uttar Pradesh},
            country={India}}
\affiliation[second]{organization={Institute for Data Science, FHNW},
            addressline={Bahnhofstrasse 6}, 
            city={Windisch},
            postcode={5210}, 
            country={Switzerland}}

\author[second]{Simon Felix}

\author[third]{Luis Fernando Machado Poletti Valle}
\affiliation[third]{organization={Institute for Particle Physics and Astrophysics, ETH Zurich},
            addressline={Wolfgang-Pauli-Strasse 27}, 
            city={Zurich},
            postcode={8093}, 
            country={Switzerland}}

\author[second]{Vincenzo Timmel}
\author[second]{Lukas Gehrig}
\author[second]{Andreas Wassmer}

\affiliation[fourth]{organization={Laboratoire d’Astrophysique, Ecole Polytechnique Federale de Lausanne EPFL}, 
            adressline={Observatoire de Sauverny},
            city={Versoix},
            postcode={1290},
            country={Switzerland}}
            
\author[third]{Jennifer Studer}
\author[third]{Pascal Hitz}
\author[second]{Filip Schramka}
\author[third,fourth]{Michele Bianco}
\author[third]{Devin Crichton}
\author[third]{Marta Spinelli}

\author[second]{Andr\'e Csillaghy}
\author[second]{Stefan K\"{o}gel}
\author[third]{Alexandre R\'efr\'egier}

\begin{abstract}
\karabo{} is a versatile Python-based software framework simplifying research with radio astronomy data. It bundles existing software packages into a coherent whole to improve the ease of use of its components. \karabo{} includes useful abstractions, like strategies to scale and parallelize typical workloads or science-specific Python modules. The framework includes functionality to access datasets and mock observations to study the Square Kilometer Array (SKA) instruments and their expected accuracy. SKA will address problems in a wide range of fields of astronomy. We demonstrate the application of \karabo{} to some of the SKA science cases from HI intensity mapping, mock radio surveys, radio source detection, the epoch of re-ionisation and heliophysics. We discuss the capabilities and challenges of simulating large radio datasets in the context of SKA. 
\end{abstract}




\end{frontmatter}




\section{Introduction}
 \label{sec:intro}

Radio observations are crucial for the understanding of the universe.
They provide a unique window to the universe, and the upcoming world's largest radio interferometer, the Square Kilometer Array (SKA), will provide unprecedented radio data quality to enable data-driven discoveries. However, in any data-driven discovery, the simulations of radio observations play a crucial role in providing a comprehensive picture in most radio science cases. They quantify the instrumental artefacts and relevance of the desired radio signal and debug the data pipelines.  

SKA is currently under construction \citep{dewdney2009square}. It consists of two instruments, SKA-Low and SKA-Mid, which cover the 80-350 MHz and 350 MHz-30 GHz frequency range, respectively. SKA will enable a vast range of science cases \citep{Braun2015aska}, with unique sets of scientific and data requirements, hence a challenge. 
Most SKA measurements will be distributed as 3D data cubes (a 2D sky measured in multiple frequencies). In rare cases, visibilities and calibration solutions will be made available as well. This presents a challenge in debugging the data signal chains. The large quantity of data presents challenges not only to data pipelines but also to data handling and analysis at the user's end. The users must be able to use their own pieces of code for analysis. The analysis tools must be user-friendly enough to appeal to a diverse range of users with different expertise. Mock radio data generation is an integral and common part of analysis tools. 

We have developed \textit{\karabo{}}\footnote{https://github.com/i4Ds/Karabo-Pipeline}, a user-friendly and versatile radio data simulator. The primary motivator of \karabo{} is SKA, allowing mock simulations from stations and antennas for SKA-low and SKA-mid. We utilise \oskar{}\footnote{https://github.com/OxfordSKA/OSKAR} and \rascil{}\footnote{https://gitlab.com/ska-telescope/external/rascil} packages as simulation engines for SKA-low and SKA-mid, respectively.
In \karabo{}, the distribution of existing radio astronomy software packages is made interoperable through common Python interfaces. Individual functions from existing packages can be combined in a number of ways to build radio-astronomy data processing pipelines for a wide variety of use cases. \karabo{} provides easy access to several real-world and synthetic datasets. In the following chapters, we outline the \karabo{} architecture and design choices and present science cases using \karabo{}.

The scale of the SKA data products exceeds that of all previous instruments \citep{quinn2015delivering}. To handle the enormous datasets, novel methods have to be developed, and existing algorithms must be adapted. Some of these novel methods might be adapted from other domains dealing with large datasets. For example, machine learning methods can be used to localize sources without imaging \citep{taran2023challenging} or for fast image deconvolution \citep{drozdova2024radio}. One of our goals is to make it easier for non-domain experts to work with radio astronomy data, algorithms, software, and challenges.

\karabo{} makes it easy to install and use all relevant software packages. Users can build intricate pipelines in just a few lines of Python code. In addition to SKA, \karabo{} also supports various existing telescopes, including SKA-precursors and pathfinders.

\subsection{Related Work}

CASA \citep{bean2022casa} served as inspiration for \karabo{}, as it follows similar ideals. It bundles many interoperable algorithms in a user-friendly package. Compared to \karabo{}, CASA is more opinionated in its data models, and more geared towards domain experts. Having a single, standardized data model improves usability and consistency and simplifies the implementation, but it also comes with its own challenges. For example, the reliance on the Measurement Set visibility format limits scalability \citep{perkins2021dask}.

\begin{figure*}
    \centering
    \resizebox{1.8\columnwidth}{!}{
        \includegraphics{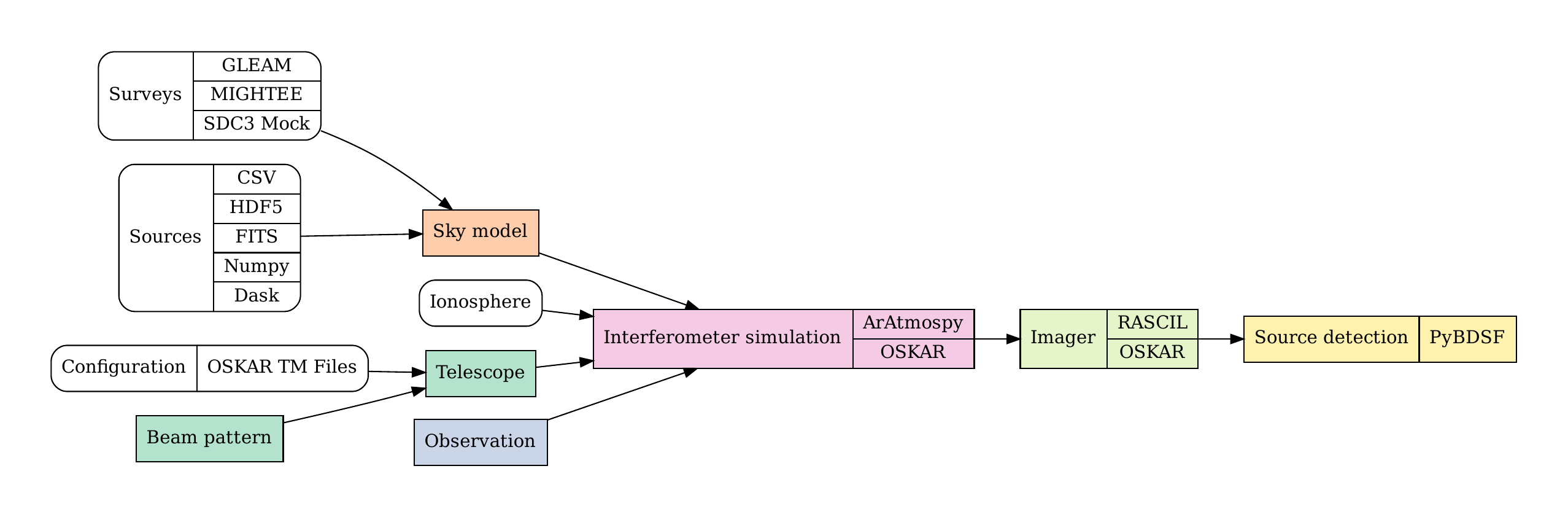}
    }
    \caption{This flowchart shows \karabo{}'s major components. Data is represented by rounded rectangles, blocks are components. The \emph{Interferometer simulation} simulates visibilities for a given sky model, telescope configuration and observation parameters. The \emph{Imager} produces images from visibilities, which can be analyzed by the \emph{Source detection} component. We assess the reliability of all components by comparing the detected sources with the sources in the sky model.}
    \label{fig:components}
\end{figure*}

\label{sec:architecture}
We wrap existing software packages with common Python interfaces for interoperability. The interfaces provide a common abstraction for different implementations, e.g., imagers. Users are free to combine the packages into complex data processing pipelines, as shown in Fig. \ref{fig:components}. The following chapters describe the fundamental components of \karabo{}.

\subsection{SkyModel}
The \skymodel{} entity defines the sources in the sky, which can be observed in later stages. We define a sky as a combination of point sources and extended sources. The minimum information needed per source is at least Stokes I/Q/U/V, but \karabo{} can manage additional information, such as Full-Width at Half Maximum (FWHM) or spectral index. We use Jansky as a common unit for the Stokes parameters or input flux density. A \skymodel{} can be created from code, loaded from well-known catalogues like GLEAM \cite{hurley2017galactic} and MALS \cite{gupta2017MeerKAT}, or read from \texttt{.h5} or catalog files. An example of a small catalog file is shown in Fig. \ref{fig:sky_model1}. Users can further visualize a \skymodel or filter sources by frequency, flux and position. The supported functions should feel familiar to users coming from \oskar{} or \rascil{}.

Not every interferometer simulation supports both extended sources or a large number of point sources. We convert between those representations as required, e.g. extended sources may be approximated by a large number of point sources. Similarly, \karabo{} may bucket continuous spectra into fixed frequency bins when required.

\begin{figure}[b]
    \centering
        \begin{lstlisting}
# Required columns:
# =================
# RA(deg), Dec(deg), I(Jy)
#
# Optional columns:
# =================
# Q(Jy), U(Jy), V(Jy), freq0(Hz), spectral index, rotation measure,
#    FWHM major (arcsec), FWHM minor (arcsec), position angle (deg)
#
#
# Two fully-specified sources
0.0 70.0 1.1 0.0 0.0 0.0 100e6 -0.7 0.0 200.0 150.0  23.0
0.0 71.2 2.3 1.0 0.0 0.0 100e6 -0.7 0.0  90.0  40.0 -10.0

# A source with only Stokes I (other attributes take default values)
0.1 69.8 1.0
\end{lstlisting}
    \caption{ Input format of source catalogues for  \oskar{} and \karabo{}. A catalogue file consists of rows equivalent to the number of sources and 13 columns. There are three columns minimum required specifying R.A., Dec and Stokes I brightness of the source.}
    \label{fig:sky_model1}
\end{figure}

\begin{figure}
    \centering
    \begin{lstlisting}[language=Python]
s = SkyModel.get_GLEAM_Sky()
s.explore_sky([250, -80], s=0.1)\end{lstlisting}
    \resizebox{\columnwidth}{!}{
        \includegraphics{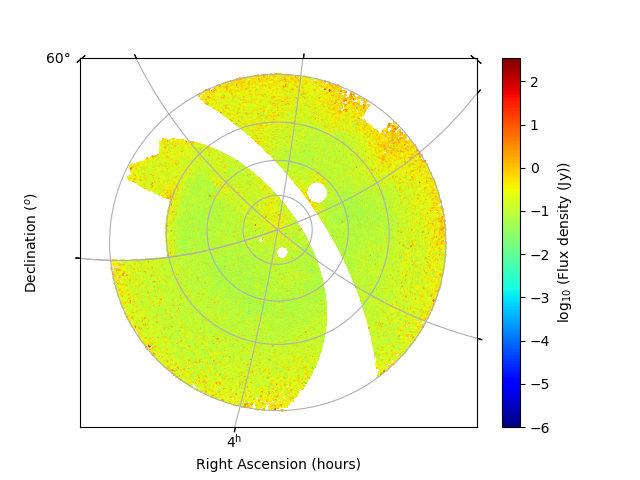}
    }\\
    (A) GLEAM Sky \\
        \resizebox{\columnwidth}{!}{
        \includegraphics{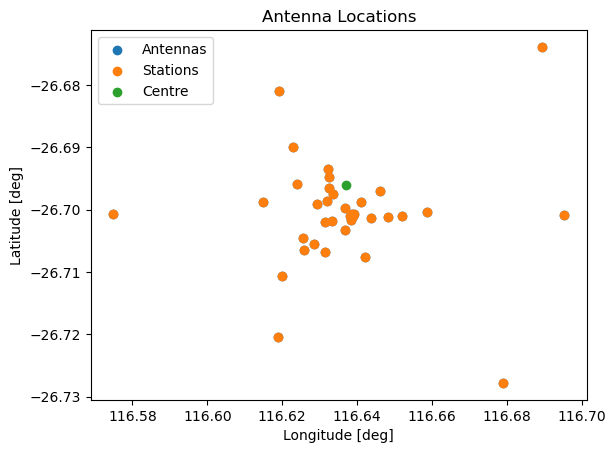}
    }\\
    (B) Example ASKAP telescope configuration
    \caption{ Top listing shows the \karabo{}'s python commands. Loading and visualising the GLEAM survey catalog requires a single line of code each. \karabo{} automatically downloads and caches catalogs on first use. Panel A: The GLEAM catalogue shows all the sources in all frequency ranges. Note that the missing stripe in the middle corresponds to the Milky way's disk. Panel B: An example of the antenna configuration for the ASKAP telescope, generated using \karabo{}'s \telescope{} class.}
    \label{fig:gleam-askap}
\end{figure}

\subsection{Telescope}
A \telescope{} is defined by the locations of its data-collecting components, which can be dishes (e.g. SKA-Mid) or stations composed of multiple antennas (e.g. SKA-Low). The positioning of the dishes/stations is determined according to the science cases \citep[e.g.][for HI studies]{Patra2019}. \karabo{} contains configurations for many radio telescopes (e.g. MWA, Very Large Array (VLA), MeerKAT), and allows users to experiment with entirely new telescope configurations. Telescope data and functionality can be accessed via the \telescope{} class, which supports \texttt{.tm} file formats, as defined by \oskar{}, see \ref{sec:interoperability}). After selecting a telescope configuration, users can further modify details with baseline filtering, and visualize the telescope configuration, as shown in Fig. \ref{fig:gleam-askap} (B).

\subsection{Observation and InterferometerSimulation}
\karabo{} supports two observation modes: short-time integration observations up to 24 hours and long-time integration observations, where observations take place over multiple days. The former supports snapshot radio observations too. \karabo{} automatically runs multiple simulations in parallel on single machines or on multi-node clusters, and combines results as needed. 

Currently, \karabo{} computes visibilities from \skymodel{}s with one of two software packages, \oskar{} or \rascil{}. \oskar{} \citep{mort2010oskar} is a package designed for the simulation of aperture array telescopes, such as SKA-Low, with GPU acceleration in mind. \rascil{} is part of the SKA Science Data Processor development and has been designed with a focus on SKA-Mid applications by simulating telescope dishes instead of the station/antenna configuration present in SKA-Low. 

Accurate simulations of beam patterns are critical for the SKA-Low and SKA-High telescopes. This poses an interesting problem: \oskar{} implements simulation of station beams using dipole configuration, while \rascil{} implements beam correction to the sky model, i.e. mimicking dishes. We build a best-of-both-worlds beam class, which automatically picks the appropriate simulation based on frequency. Of course, users are free to override this decision when they require more control.

\subsection{Imager}

The \imager{} class computes images from measured or simulated visibilities. The current version supports three imagers: \oskar{}, \rascil{} and \wsclean{} \citep{offringa2014wsclean}. The \oskar{} imager can only make dirty images, while \rascil{} imager can make clean images with the CLEAN algorithm. \wsclean{} is a generic wide-field imager with support for large data volumes. We plan to add more imagers in the future.

The SKA-Low and SKA-Mid observation strategies will involve measuring data from multiple pointings in the sky and combining the results into a single mosaic. \karabo{} supports mosaicing through the \imagemosaicker{} class, which combines a set of \image{} objects into one \image{} mosaic. \karabo{} reprojects the images appropriately with the help of Astropy's \citep{astropy:2022} reproject package.



\section{Design Priorities}\label{sec:design}

\subsection{Making Software Packages Interoperable}\label{sec:interoperability}
Some of the different software packages that make up \karabo{} use different data representations. Additionally, the packages often support different configuration parameters. \karabo{} converts between data formats on-the-fly, as required, and exposes a common set of standardized configuration options. This support for multiple packages allows users to swap and compare packages with minimal configuration and code changes.

\subsection{Scalability and Parallelization}
Next-generation radio-astronomy telescopes produce data at unprecedented rates, leading to challenges in analysis and processing. Simulations of these instruments require significant computational resources. The majority of astronomical data processing will be performed on supercomputer clusters with hardware accelerators.

\karabo{} supports parallelization and GPU-processing out of the box. \karabo{} automatically detects the available GPU hardware and whether it runs on a multi-node cluster infrastructure. It automatically configures the software in a processing pipeline to make maximum use of the available hardware. Parallelization to multiple nodes, as well as aggregation of the data products, relies on data-partitioning with Dask \citep{dask2015}. \karabo{} also supports out-of-core dataset sizes with streaming. We plan to enhance the scalability of \karabo{} further in the future.

\begin{figure}[ht]
    \centering
    \resizebox{\columnwidth}{!}{
        \includegraphics{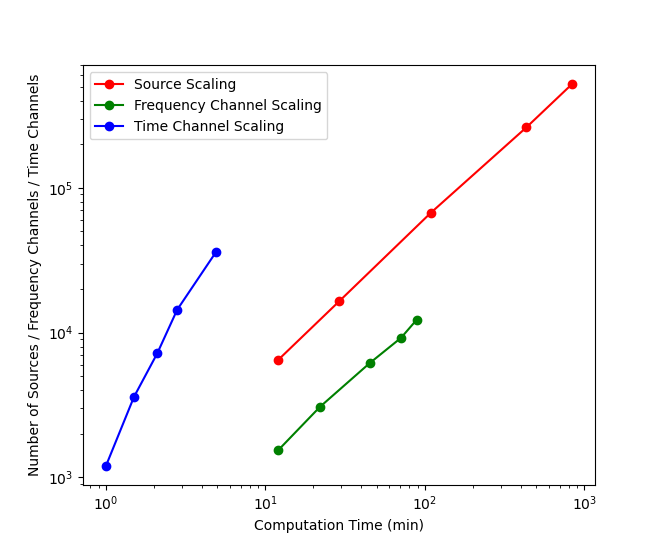}
    }
    \caption{Computation time for different run settings showing the scaling with number of sources (red), frequency channels (green) and time channels (blue).}
    \label{fig:ct_scaling}
\end{figure}

\subsection{Software Development and Deployment}
We rely on automated processes and tooling to increase the reliability of \karabo{}. For every code change we make, our CI/CD infrastructure automatically runs type checkers, linters and test suites.

We found that installing radio-astronomy software packages can be difficult for a variety of reasons: mismatched library versions, incompatible compilers, slow installation procedures, incompatible operating systems, undocumented or outdated dependencies or packages using incompatible file formats. To avoid many of these potential problems \karabo{} installs pre-compiled binary packages through the Conda ecosystem. \karabo{} installs native libraries without requiring system-wide changes. We also provide container images for Docker or Singularity as an alternative installation option.

\subsection{Runtime scaling in different radio data dimensions}

Radio data is generally arranged in multidimensional arrays of polarization, baselines, time channels, frequency channels, and the number of sources. For a given telescope configuration, the frequency and time integrations of the observations are designed to dictate computational requirements. In simulations, the number of sources fed into the interferometer module also plays an important role. For a meaningful deployment of the pipeline, one needs to assess the computational requirements of the simulation run. Although the exact speeds and run time of the simulation will depend on the computing infrastructure, a rough notion of the run time is important. We measured the run time for 15 simulation runs for the same observations with varying time, frequency channels and number of sources, i.e. 5 runs each. To investigate the time scaling, the number of sources and frequency were fixed. Fig. \ref{fig:ct_scaling} shows the scaling of runtimes, and we note that the number of sources (red) impacts the runtime most, while time channels (blue) have the least impact. Therefore, simulation runs with a large number of sources and frequency channels are the most expensive, while runs with a smaller number of sources and long-time integrations are relatively cheaper.

\section{Science Use Cases}\label{sec:science}
SKA will enable a wide range of sciences \citep{braun2015advancing}, thus \karabo{} should also support a wide range of use cases. We started by reproducing the well-known MeerKAT International GHz Tiered Extragalactic Exploration (MIGHTEE) survey. For this, we simulated observations of the MIGHTEE point source catalogue. Similarly, we also ran MWA simulations of the GLEAM point source catalogue. The other science cases we tested are radio point source simulation and detection in images, simulating sky signals and various foregrounds for EoR, and simulating the extended solar disk for heliophysics applications. In addition, we also developed a point subtraction tool for foreground removal in the SKA Science Data Challenge 3a\footnote{\url{https://sdc3.skao.int/}}.

\subsection{HI Intensity Mapping}

Neutral hydrogen (HI) intensity mapping (IM) is an important observational technique to study the distribution of dark matter, constrain the dark energy equation of state and associated science questions \citep{Santos2015aska}. The observation of the 21 cm line emission caused by its hyperfine transitions makes it possible to tomographically probe large, cosmological volumes. HI is a tracer of the underlying dark matter distribution in the post-EoR Universe ($z < 6$), which helps to reconstruct the dark matter density fields. Therefore, large-scale radio surveys like SKA are a useful tool for cosmology to study the dark matter density fields.

\begin{figure*}
    \centering
\begin{lstlisting}[language=Python]
from karabo.imaging.image import ImageMosaicker
mi = ImageMosaicker()
restored_mosaicked = mi.mosaic(restored_cuts)
restored_mosaicked[0].plot()
\end{lstlisting}
    \begin{tabular}{ccc}
        \resizebox{0.7\columnwidth}{!}{ \includegraphics{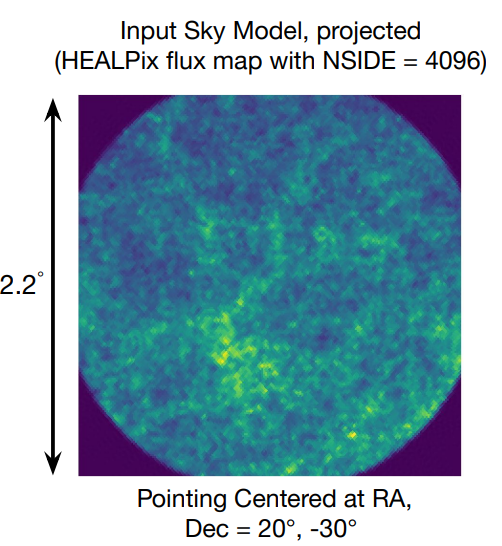}}  &
       \resizebox{0.7\columnwidth}{!}{  \includegraphics{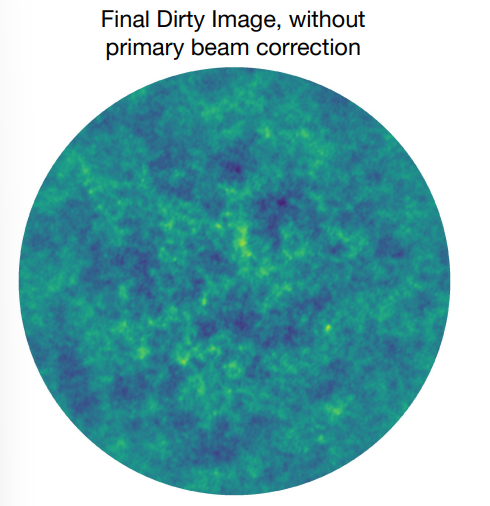}} &
       \resizebox{0.7\columnwidth}{!}{  \includegraphics{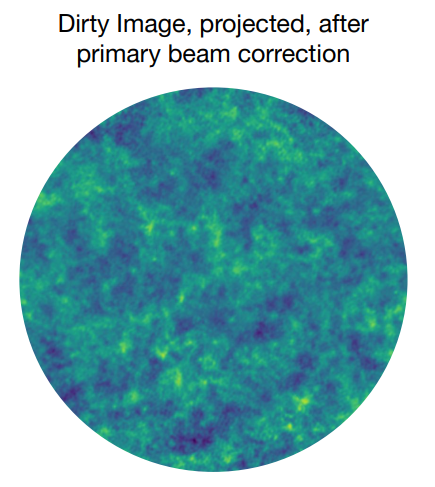}}
        \\
        (A) Skymodel Image  & (B) Dirty image & (C) PB corrected Dirty image  \\
    \end{tabular}
   \resizebox{1.2\columnwidth}{!}{  \includegraphics{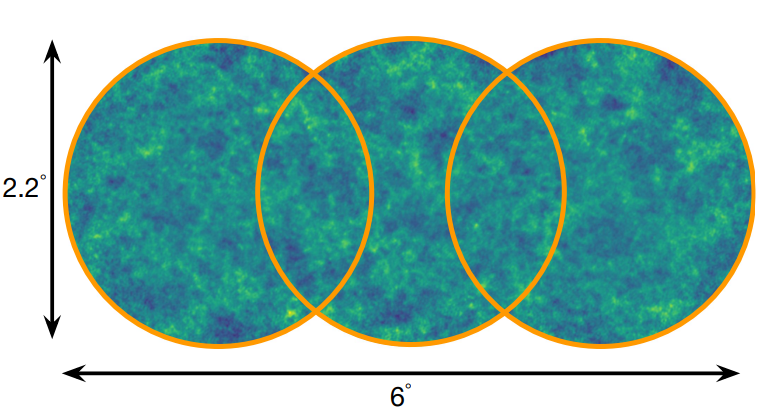}}\\
   (C) Mosaic for 3 pointings
    \caption{Panel A: A cutout from the input sky image obtained using the dark matter halo simulations from Pinocchio software. Panel B: Dirty image before the primary beam correction. Note the faint emission near the boundaries. Panel C: Dirty image with primary beam correction. We have taken a MeerKAT primary beam size with a Gaussian profile for the demonstration. Panel D: Mosaic image obtained after sticking three images, of which the two images are with 0.5 deg in pointing apart w.r.t to the center of the skymodel. The top listing shows the Python commands to call the mosaicing function and construct the mosaicc from a list of images (\texttt{restored\_cuts}).}
    \label{fig:im}
\end{figure*}

Dark matter simulations based on hydrodynamical models can be used to estimate the mass of the HI distribution and the corresponding brightness temperature distribution \citep{Hamsa2022MNRAS}. Such models can be passed through \karabo{} to create mock HI-intensity maps from a radio telescope. 
For demonstration, we have used state-of-the-art dark matter simulations using the PINpointing Orbit Crossing-Collapsed HIerarchical Objects (PINOCCHIO) algorithm \citep{Manaco2013MNRAS}, \citep{pinocchio2002} and \citep{pinocchio2017}. We chose a $2048^3$ box, 500 Mpc/h, and $0.77<z<1.03$ corresponding to 700-800 MHz for the dark matter run. For the simulation, we split the map into equally spaced frequency bins, which defines the redshift resolution of our HI IM simulation. We then simulate the HI IM for every frequency bin map in the chosen telescope and observation configuration. For that, we convert the redshift to the corresponding observed frequency. We tested \karabo{} simulations with the SKA-Mid and MeerKAT configuration (Fig. \ref{fig:im}). Since we are interested in the large-scale distribution of HI, we need several pointings of the sky. \karabo{} contains a mosaicing routine with which one can simulate a larger area of the sky.

Fig. \ref{fig:im} (A) shows the flux density of the input sky restricted to the same area as the simulated pointing (Fig. \ref{fig:im} (B)) and plotted with \healpix{} \footnote{\url{http://healpix.sourceforge.net}}. Using \karabo{}'s HDF5 file interface, we import the map and build a sky model with the centre at R.A. $20^o$, and DEC $-30^o$ and a diameter of $10^o$. The simulation was done for the MeerKAT configuration between 700 MHz and 800 MHz with a bandwidth of 5 MHz, and with a primary beam's full width at half maxima (FWHM) of 1.8$^o$. Fig. \ref{fig:im} (B) shows the dirty image, followed by the primary beam corrected image (C) and a mosaic of 3 pointings (D). Common emission features between the dirty image and the input sky model can be observed (A). The resulting mosaic has a size of $2^o \times 6^o$. The dirty image and mosaic shown are added up along the redshift axis. However, the simulation also provides three-dimensional images in the chosen redshift resolution.

\begin{figure*}
    \centering
\begin{lstlisting}[language=Python]
sky = SkyModel()
mightee_skymodel = SkyModel.get_MIGHTEE_Sky()
-----------------------------------------------
sky = SkyModel()
gleam_skymodel = SkyModel.get_GLEAM_Sky()
\end{lstlisting}  
    \begin{tabular}{cc}
        \resizebox{0.9\columnwidth}{!}{ \includegraphics{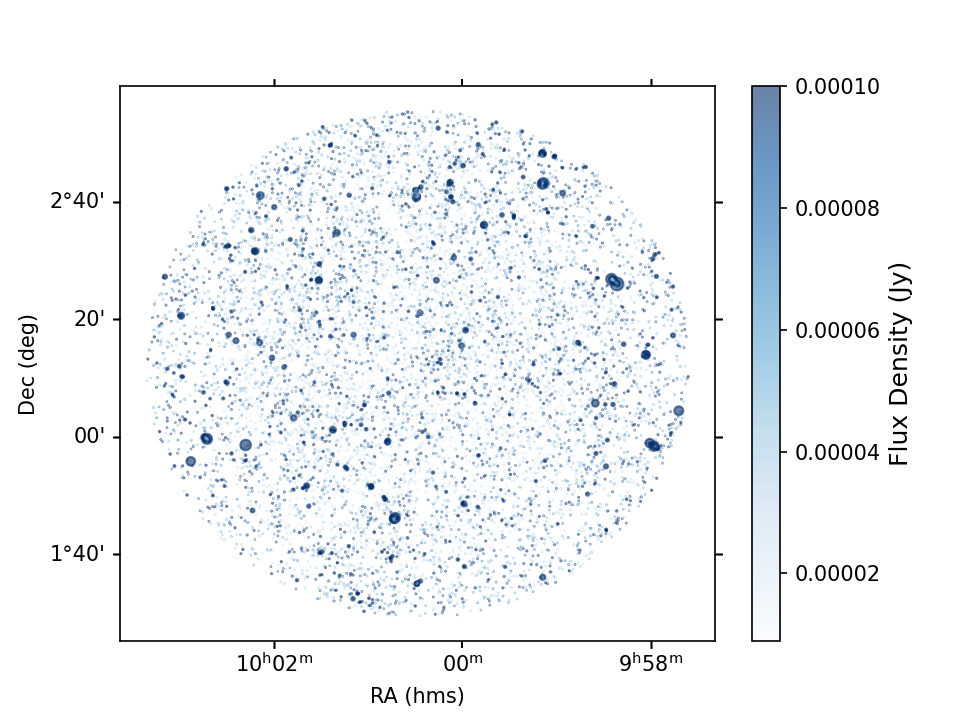}}  &
       \resizebox{\columnwidth}{!}{  \includegraphics{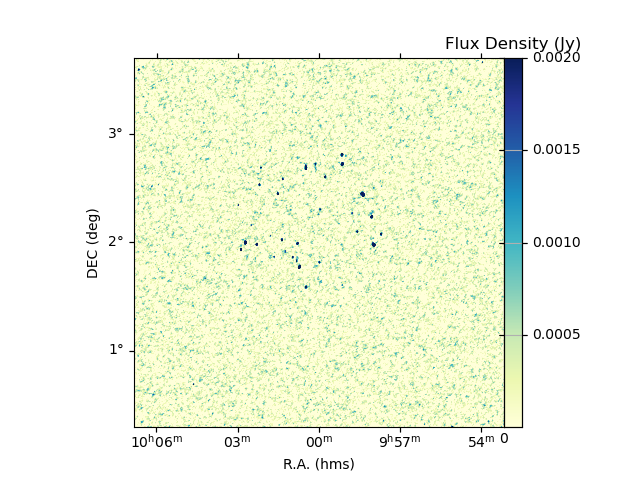}}
        \\
        (A) Input MIGHTEE skymodel & (B) Mock image from MeerKAT \\
        \resizebox{\columnwidth}{!}{ \includegraphics{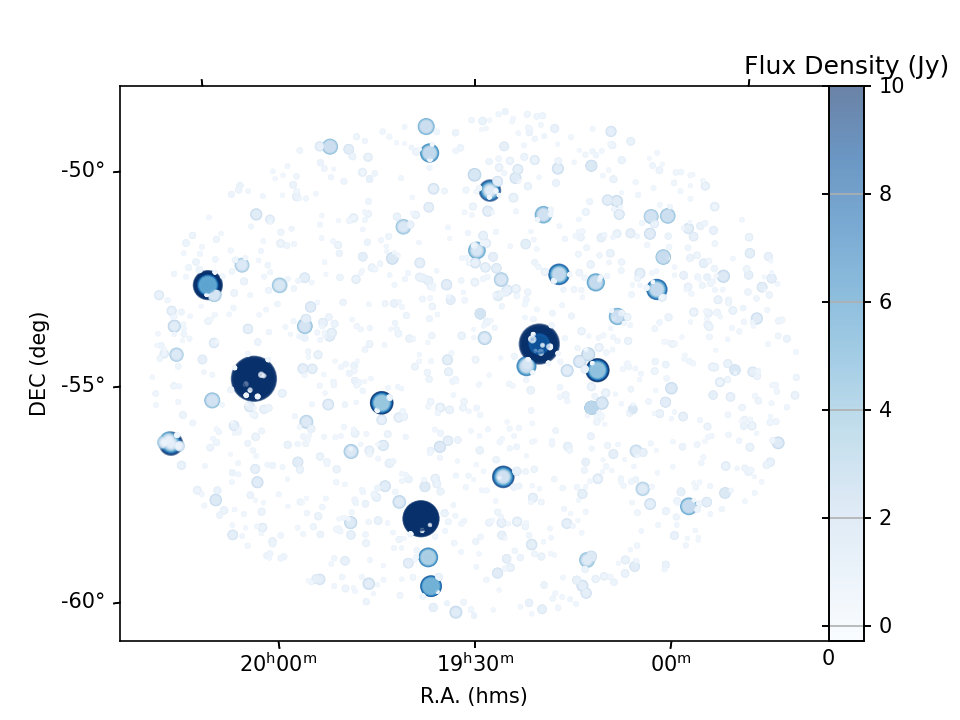}}  &
       \resizebox{\columnwidth}{!}{  \includegraphics{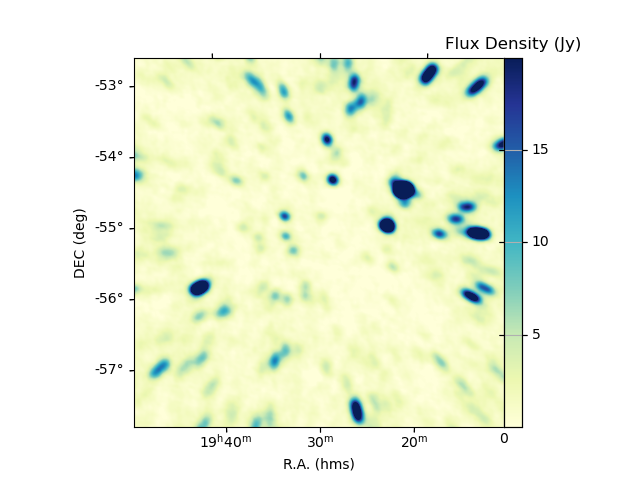}}
        \\
        (C) Input GLEAM skymodel & (D) Mock image from MWA \\
    \end{tabular}
    \caption{Example simulated images from a section of the MIGHTEE and GLEAM point source catalog for demonstration. Panel A: Input skymodel from the mightee field-of-view. The colour and size denote the flux density of the radio source. Panel B: Simulated images from Karabo for a 5-hour-long time integration from MeerKAT telescope. Panel C: Input skymodel from the GLEAM point source catalog. Panel D: Simulated image of the GLEAM skymodel from MWA telescope with 1-minute time integrations. We can see that the bright radio sources in the images can be matched with their respective input images. The top listing shows the Python commands to call MIGHTEE and GLEAM catalogs.}
    \label{fig:mock}
\end{figure*}

\subsection{Mock Continuum Survey}
Radio continuum observations play an essential role in various science cases, for example relating to galaxy studies or cosmology. A radio continuum is a continuous emission in frequency space that arises from multiple types of objects, like active galactic nuclei (AGN), radio lobes, star-forming regions, diffuse thermal and gyrosynchroton emission, etc. This section demonstrates simple mock data simulation for a limited MeerKAT International GHz Tiered Extragalactic Exploration \citep[][MIGHTEE]{Jarvis2016} and GaLactic and Extragalactic All-sky MWA survey \citep[][GLEAM]{Wayth2015} point source catalogs. These were observed by MeerKAT and MWA, the SKA precursors.

%

\subsubsection{Mock MIGHTEE catalog image}

The MIGHTEE survey by the MeerKAT telescope is one of the modern surveys covering a 20 $deg^2$ field-of-view (FoV) over the COSMOS, E-CDFS, ELASIS-S1 and XMM-Newton Large Scale Structure field deep fields at SKA-mid frequencies. 
Using the point source catalogue from MIGHTEE over a $3^{\circ}$ COSMOS FoV, we simulate observations from the MeerKAT telescope configuration, which has 64 antennas. The pointing centre was at R.A. = $150^{\circ}$, DEC = $2^{\circ}$. Within the FoV, the catalogue has $\sim9896$ sources with a distribution in the flux densities with their respective spectral indices. We demonstrate a snapshot simulation, which consists of an 8-second time integration and 0.2 MHz channel bandwidth integration. Measurement sets were created for the total observation was done between 856 MHz and 1675.2 MHz with 4096 frequency channels.
The clean images from the measurement set of the observation were created, followed by imaging using \wsclean{} deconvolution. The point sources from the catalogue and the simulated deconvolved image are shown in Fig. \ref{fig:mock} (A) \& (B). 

\subsubsection{Mock-GLEAM catalog image}
In SKA-low frequencies, the MWA's GLEAM is the most comprehensive wide-field survey covering 24,831 $deg^2$. This large FoV was achieved by drift-scanning using MWA's fine 40 kHz frequency channel width and 0.5 sec time resolution. GLEAM contains more than 0.3 million components. 
We use a subset of these sources in our simulation to demonstrate the mock data creation at SKA-low frequencies using MWA phase-I configuration. We use R.A. and DEC of the pointing at $292.5^{\circ}$ and $-55^{\circ}$, respectively. We limit the sky model FoV to $6^{\circ}$ around the pointing centre, with 23600 radio sources. 
We performed a snapshot observation for demonstration, covering an observation duration of 1 minute with 0.5-second time integration and 40 kHz channel bandwidth. 

Fig. \ref{fig:mock} (C) shows the input GLEAM sky to \karabo{}, while panel (D) shows the simulated snapshot images from the MWA telescope. As expected, we notice the brightest regions in the simulated images corresponding to the input model (panels A \& C).

\subsection{Source Detection}
\begin{figure}
    \centering
        \begin{lstlisting}[language=Python]
from karabo.sourcedetection.evaluation import SourceDetectionEvaluation
sde = SourceDetectionEvaluation( sky=sky, ground_truth=ground_truth,
        assignments=assignments, sky_idxs=sky_idxs,
        source_detection=detection_result,)
sde.plot()
\end{lstlisting}
       \resizebox{0.7\columnwidth}{!}{  \includegraphics{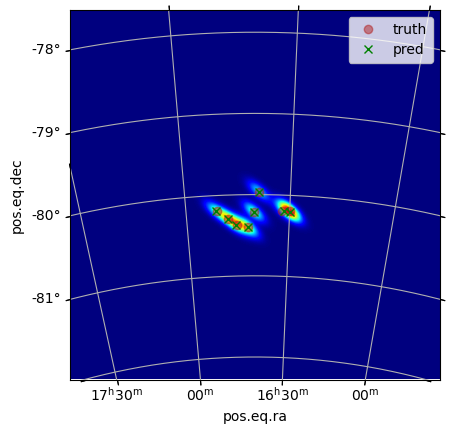}}
    \caption{\karabo{} was used to simulate the ASKAP observation of sources from the GLEAM catalog. WSCLEAN was used for imaging, and PyBDSF to detect sources. \karabo{} automatically matches the detected sources with those from the catalog, and computes the accuracy of source detection. Top listing shows the function for source detection evaluation from image (sky), and plotting the result.}
    \label{fig:source-detection}
\end{figure}

Many problems, ranging from galactic and extragalactic studies to cosmology, desire the detection of faint radio point sources. All science cases that build on source statistics and source characterisation are based on source detection. The objective of source detection is to detect radio sources with a given significance level and for each source determine additional parameters like size, flux densities, and spectral index. 
\karabo{} performs source detection with PyBDSF \citep{mohan2015pybdsf}. This is useful by itself, but \karabo{} can also be used to evaluate the accuracy of source detection algorithms, thanks to built-in evaluation metrics like precision and recall. \karabo{} can compare the found sources with known sources from a catalog. A demonstration of the source detection from \karabo{} is shown in Fig. \ref{fig:source-detection}. Here, we simulated 8 point sources to be detected. Using the ASKAP telescope configuration (Fig. \ref{fig:gleam-askap} (B)), we simulate visibilities at starting frequency 100 MHz, channel bandwidth 1 MHz, 16 frequency channels and 24 time channels. We used the RASCIL imager for the dirty image and WSCLEAN for a cleaned version. The cleaned image was passed to PYBDSF, and source detection was performed (Fig. \ref{fig:source-detection}).

\subsection{Epoch of Reionisation}

\begin{figure*}
    \centering
\resizebox{2.00\columnwidth}{!}{ \includegraphics{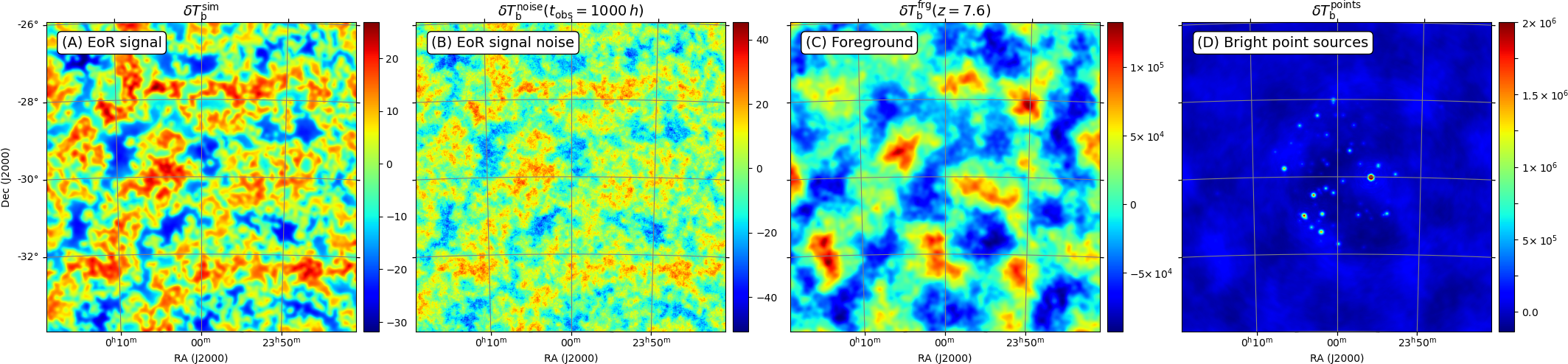}}
    \caption{EoR simulation products made from \karabo{}. Panel A: Brightness temperature fluctuation of an EoR signal. This skymodel was obtained from the tools21cm package. Panel B: Gaussian noise on the EoR signal for a hypothetical synthesis of 1000 h radio observation. Panel C: Foreground emission at the level of $\sim$1000 K. Panel D: Bright point sources adding to the complexity to the foreground. Note that the EoR signal is four orders of magnitude smaller than the bright point sources in the field of view.}
    \label{fig:eor}
\end{figure*}


After the big bang, the recombination of hydrogen atoms resulted in an 21cm hydrogen line emission. Later, as the first stars formed in our universe, the resulting heating changed the absorption profile of this emission. The resulting subtle changes with redshift or observing frequency is called the Epoch of Reionization (EoR) signal \citep[EoR][]{Zaroubi2013}. 

EoR measurements are challenging due to the weak and extended nature of the signal, the presence of strong and diverse foregrounds, ranging from nearby strong point sources, extended galactic synchrotron emission, weak extra-galactic point source populations, and others. \karabo{} can simulate EoR signals with various foreground signals making it a great tool for EoR simulation studies. Major foreground signals that can be used are GLEAM catalogue point sources, extended Haslam synchrotron maps, and others. The tools21cm package \citep{giri2020tools21cm} is integrated into \karabo{} to simulate the EoR signal itself. The combined data offers an exciting glimpse into the future EoR studies that SKA makes possible.

Fig. \ref{fig:eor} (A) and (B) shows EoR signal and signal noise simulated from tools21cm modules integrated into the \karabo{} package. As a part of SKA Data Challenge 3a, the SKACH team used \karabo{} simulations to study EoR detection and the role of the foreground in the images seen through the SKA-mid configuration \citep{Bonaldi2025,Bianco2024}. The primary goal of the data challenge was to remove the foreground and recover the EoR signal. Panels (C) and (D) show the galactic foreground and bright point sources, respectively. The bright point sources produce strong sidelobe artefacts which must be removed to recover the faint EoR signal. The point source removal is more effective in visibility space than in the image plane \citep{Sharma2022ApJ}. We integrated such a visibility point source removal algorithm into \karabo{}.



\subsection{Heliophysics}
The Sun is the nearest and brightest radio source, and most important to human space-based activity. The energetic particle beams originating in the corona are the driver for space weather and need to be understood in terms of their time and frequency variability \citep[e.g.][]{Harra2021}. Radio observations of the sun increase our understanding of the corona, the outermost layer of the sun, and the interplanetary medium. The high sensitivity of the SKA would provide solar images at an unprecedented time and frequency resolution \citep{Nindoas2019,Oberoi2023}. Using \karabo{}, we can estimate the expected levels of detail of SKA images for different activity levels of the Sun.

\begin{figure*}
    \centering
    \begin{tabular}{ccc}
        \resizebox{0.65\columnwidth}{!}{ \includegraphics{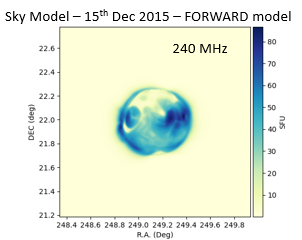}}  &
       \resizebox{0.75\columnwidth}{!}{  \includegraphics{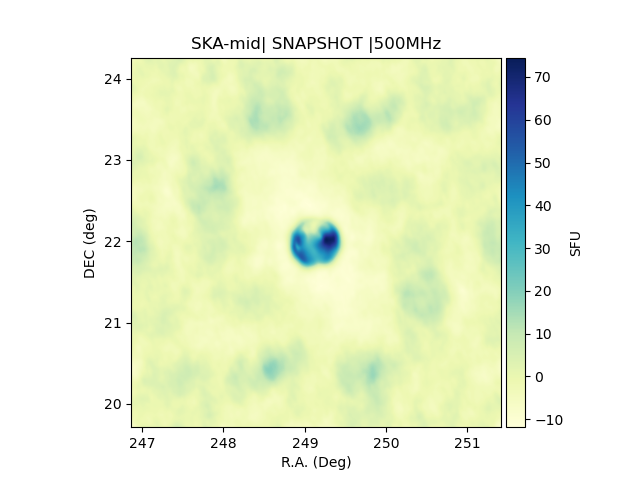}} &
       \resizebox{0.75\columnwidth}{!}{  \includegraphics{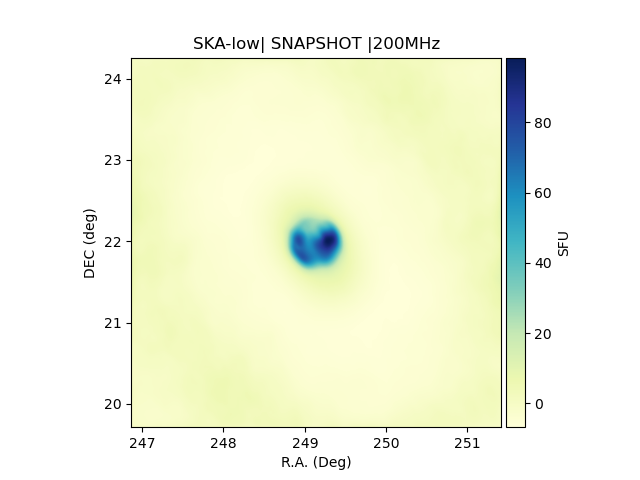}}
        \\
        (A) FORWARD solar skymodel & (B) SKA-mid image & SKA-low image \\
    \end{tabular}
    \caption{Simulation of the quiet Sun disk. Panel A: The skymodel of the flux densities was obtained from the FORWARD model by utilizing the empirical MHD model of the solar atmosphere at 240 MHz. Panel B: Simulated snapshot image of the solar disk from SKA-mid at 500 MHz. Panel C: Simulated snapshot image of the solar disk from SKA-low at 200 MHz. Note that the flux densities have been scaled with a spectral index of 2.}
    \label{fig:helio}
\end{figure*}

Fig. \ref{fig:helio} shows an example of the simulation of a solar disk for a snapshot (8-sec time integration). The sky model was obtained by the FORWARD model \citep{Gibson2016}, which simulates the radio emission from thermal bremsstrahlung based on realistic magnetohydrodynamic models from magnetic field maps. We note that the SKA-mid and SKA-low simulated images have good dynamic range. The SKA-low image is more diffuse than SKA-mid image at high frequencies. Many types of such solar simulation studies can be done using \karabo{}, like simulation of spatio-temporal features of active regions, level of self-noise in the image, and many others. Also, the effect of sidelobes produced by the Sun in non-solar observations can be simulated.

\section{Discussion \& Summary}\label{sec:discussion}

SKA data simulations give first-hand insights into the upcoming instrument capabilities and challenges. The development and operations of SKA instrumentation are challenges in themselves \citep{Norris2011,Tao2019}. In addition, radio data analysis of a wide range of science topics is non-trivial and requires simulation support. Simulating a wide range of science cases using common software presents unique capabilities and challenges. The primary challenge is to incorporate multiple interactions of radio waves from celestial radio sources with the telescope configurations, radio frequency environment, ionosphere and instrumental systematics, followed by producing realistic 3-D data cubes. Instrumental systematics is a long list of effects known in radio telescope operations that result in complex changes in the UV plane.

Karabo is an open-source software designed to simulate radio datasets for a wide range of science cases. It tackles all the mentioned challenges systematically.
In this study, we demonstrated some of the simulations of a subset of science topics, i.e. mock continuum survey, point source detection, EoR signal and simulation of radio Sun useful for cosmology, transients, EoR and heliophysics science. Some of \karabo{}'s unique capabilities are as follows:

\begin{enumerate}
    \item \karabo{}'s framework allows both SKA-low and SKA-mid science, i.e. useful for science cases requiring both telescopes in frequency and time.
    \item Many software tooling focuses on a single science case, which \karabo{} combines to support a broader scope.
    \item \karabo{} includes relevant metadata such as telescope configurations, and also data sets like sky surveys.
    \item End-to-end implementation of radio data analysis starting from sky models, processing of visibilities, to deconvolution, followed by image data analysis.
    \item \karabo{} scales from personal computers to HPC clusters and automatically makes use of the available hardware.
\end{enumerate}

As SKA-low and SKA-mid sites cover separate frequency bands, the science cases featuring the frequency band edges can be studied in a friendly way using \karabo{}. In addition, the location of the low and mid telescopes are far apart and lie in GMT+8 and GMT+2 time zones, respectively. The time difference of 6 hours allows for common time windows for simultaneous observations. \karabo{} can be used to design such observations and study realistic problems in imaging the desired field of view. 

Different fields of radio astronomy follow their software developments. The tools developed are necessary for the overall data analysis and have specific features dedicated to the radio astronomy field. The challenge originates from focused developments with limited interfaces. The interfaces in \karabo{} support multiple implementations for different science cases. In addition, \karabo{} also interfaces with different data formats. Depending on the need of the user, software integration can also occur at the data analysis end, i.e., analysing the images and data cubes as demonstrated by source detection in the section.

\karabo{} supports multiple telescope configurations, including SKA precursor and pathfinder telescopes. This can be used to mimic already working telescopes and preparing workflows involving real data from precursors and pathfinder telescopes. \karabo{} is an excellent tool for building mock data for source detection, such as studies and machine learning applications. 

\karabo{} is a continuing development aimed at systematically simulating SKA datasets in sync with the development of different array assemblies. The flexible architecture of \karabo{} also allows implementing the instrumental systematics from actual measurements from SKA's array assemblies. The inclusion of existing and custom telescope configurations, along with the inclusion of realistic effects, makes \karabo{} not only interesting for SKA simulations but a powerful, general tool to work with radio observations.

\section*{Acknowledgements}
This work was supported in part by the AstroSignals Sinergia grant No. CRII5\_193826 from the Swiss National Science Foundation. This work was supported by SERI as part of the SKACH consortium. We also thank Christoph V\"ogele in the initial stages of the project.
This work was supported by a grant from the Swiss National Supercomputing Centre (CSCS) under project ID \texttt{sk05}. 
 Some of the results in this paper have been derived using the healpy and HEALPix packages.

%

\vspace{5mm}

\software{:
    astropy \citep{astropy:2022, astropy:2018, astropy:2013},
    OSKAR \citep{mort2010oskar},
    PyBDSF \citep{mohan2015pybdsf},
    WSCLEAN \citep{offringa2014wsclean},
    tools21cm \citep{giri2020tools21cm},
    healpy \citep{Zonca2019},
    HEALPix \citep{Healpix2005}
}

\appendix



\bibliography{main}
\bibliographystyle{elsarticle-harv}





\end{document}